\documentclass[preprintnumbers,10pt,nofootinbib]{revtex4}

\usepackage{amsmath,latexsym,amssymb,amsfonts}
\usepackage[pdftex]{color,graphicx}
\usepackage{bm}

\begin{document}


\title{\textbf{Spacetime surgery for black hole fireworks}}

\author{
Wei-Chen Lin$^{a,b}$\footnote{{\tt archennlin@gmail.com}},
Dong-han Yeom$^{a,d,e,f}$\footnote{{\tt innocent.yeom@gmail.com}}, and 
Dejan Stojkovic$^{c}$\footnote{{\tt ds77@buffalo.edu}}
}

\affiliation{
$^{a}$Center for Cosmological Constant Problem, Extreme Physics Institute, Pusan National University, Busan 46241, Republic of Korea\\
$^{b}$Department of Physics, Pusan National University, Busan 46241, Republic of Korea\\
$^{c}$HEPCOS, Department of Physics, SUNY at Buffalo, Buffalo, NY 14260, USA\\
$^{d}$Department of Physics Education, Pusan National University, Busan 46241, Republic of Korea\\
$^{e}$Research Center for Dielectric and Advanced Matter Physics, Pusan National University, Busan 46241, Republic of Korea\\
$^{f}$Leung Center for Cosmology and Particle Astrophysics, National Taiwan University, Taipei 10617, Taiwan
}

\begin{abstract}
We construct an explicit model for the black hole to white hole transition (known as the black hole fireworks scenario) using the cut-and-paste technique. We model a black hole collapse using the evolution of a time-like shell in the background of the loop quantum gravity inspired metric. We then use the space-like shell analysis to construct the firework geometry. Our simple and well defined analysis removes some subtle issues that were present in the previous literature. In particular, we demonstrate that the null energy condition must be violated for the bounce. 
We also  calculate the proper time scales required for the black to white hole transition, which in any valid scenario must be shorter than the evaporation time scale. In contrast, we show that the bouncing time for the distant observer can be chosen arbitrarily, since it 
 is determined by how one cuts and pastes the spacetimes outside the event horizon, and thus does not have any obvious connection to quantum gravity effects.
\end{abstract}

\maketitle

\newpage

\tableofcontents

\section{Introduction}

The issue of formation and evaporation of a black hole is very important for understanding the nature of quantum gravity. In particular, this issue is related to the information loss problem of an evaporating black hole \cite{Hawking:1976ra}. Is there a unitary theory of quantum gravity that explains the unitary evolution of evaporating black holes? If there is, is this theory consistent with the semi-classical description \cite{Yeom:2009zp}? Will the classical singularity survive in the regime where quantum gravitational effects are dominant \cite{Bouhmadi-Lopez:2019kkt,Bogojevic:1998ma}?

It is clear that understanding the fate of the singularity is very important to obtain the complete answer to the black hole evaporation and the information loss problem. Intuitively, we may classify two ways. First, we may address this problem \textit{by introducing a wave function}, i.e., by solving the Wheeler-DeWitt equation \cite{DeWitt:1967yk}. In this approach, we need to solve the Wheeler-DeWitt equation (or some version of it) and interpret the solution in the classical background, which is sometimes a subtle problem \cite{Brahma:2021xjy,Saini:2014qpa,Greenwood:2008ht,Wang:2009ay}; for an attempt to model quantum radiation from quantum background, see \cite{Vachaspati:2007hr}. Second, we may remove the singularity \textit{by introducing an effective matter} \cite{Ayon-Beato:1999kuh,Nicolini:2005vd,Frolov:1988vj,Yeom:2008qw,Bouhmadi-Lopez:2021zwt}. As a result, one could extend the effectively classical spacetime beyond the singularity. However, we need to justify the ad hoc introduced matter from the first principles, which is usually a difficult task.

Interestingly, an approach coming from the loop quantum gravity provides a method that is in between these two ways. In that approach, one first needs to solve the Wheeler-DeWitt equation in order to obtain the physical quantum state of the singularity. Usually, it is not easy to solve the Wheeler-DeWitt equation directly. However, one may reasonably expect an effective modification of the Hamiltonian which includes loop quantum gravitational effects \cite{Bojowald:2018xxu}. With this modified Hamiltonian, one can solve a set of semi-classical equations and obtain a spacetime that includes loop quantum gravitational effects, e.g., resolution of the singularity.

A typical solution in the framework of the loop quantum gravity includes \textit{bouncing} of the collapsing object \cite{Ashtekar:2018lag,Ashtekar:2018cay}. Bouncing inside the horizon is not a very surprising scenario, except for some technical issues \cite{Bouhmadi-Lopez:2019hpp}. However, in reality, this is not easy to generalize to global spacetimes in an evaporating background. In some cases, inconsistencies may arise \cite{Hong:2022thd}. In an evaporating background, the bouncing spacetimes have to consistently connect not only inside but also outside the horizon \cite{Brahma:2021xjy,Ashtekar:2005cj}. This might be realized by cutting and pasting spacetimes, e.g. like in the Haggard-Rovelli model \cite{Haggard:2014rza}. Moreover, if we modify the interior of the black hole solution, one can obtain a bouncing model that has two horizons \cite{Han:2023wxg}. The scenario proposed in \cite{Haggard:2014rza,Han:2023wxg} is also known as the \textit{black hole fireworks}.

One can revisit the cut-and-paste technique of \cite{Haggard:2014rza} and \cite{Han:2023wxg} using the thin-shell approximation \cite{Israel:1966rt}. This spacetime surgery could explain the global spacetime of the Haggard-Rovelli model \cite{Brahma:2018cgr}. For this purpose, one needs to introduce a space-like shell and paste two space-like hypersurfaces. To do this in a self-consistent way, a space-like matter shell that violates the null energy condition and reaches asymptotic infinity is required.

In this paper, we further extend this idea to the model in \cite{Han:2023wxg}, which contains two horizons. We consider a time-like shell that describes a collapsing star interior and the dynamical formation of a black hole. In addition, we cut and paste two spacetimes to accommodate a bouncing spacetime, and also cover both the outer and inner apparent horizons. This approach is technically well-defined and hence allows a more concrete and reliable way to evaluate the transition time scale from the collapsing to the bouncing phase. Unless this time scale is sufficiently long, this process will be already excluded by astrophysical observations. 

This paper is organized as follows. In Sec.~\ref{sec:bla}, we describe the black hole bouncing model of \cite{Han:2023wxg} in which a black hole phase is followed by a white hole phase. In Sec.~\ref{sec:tim}, we consider a collapsing time-like shell and the dynamical process of black hole formation. In Sec.~\ref{sec:spa}, we consider a space-like shell that separates the black hole and white hole phases in a cut-and-paste procedure. In Sec.~\ref{sec:bou}, we discuss the bouncing time scales from the black hole and white hole phases. Finally, in Sec.~\ref{sec:dis}, we summarize our results and discuss possible future research.

\section{\label{sec:bla}Bouncing black hole model}

\begin{figure}
\begin{center}
\includegraphics[scale=0.5]{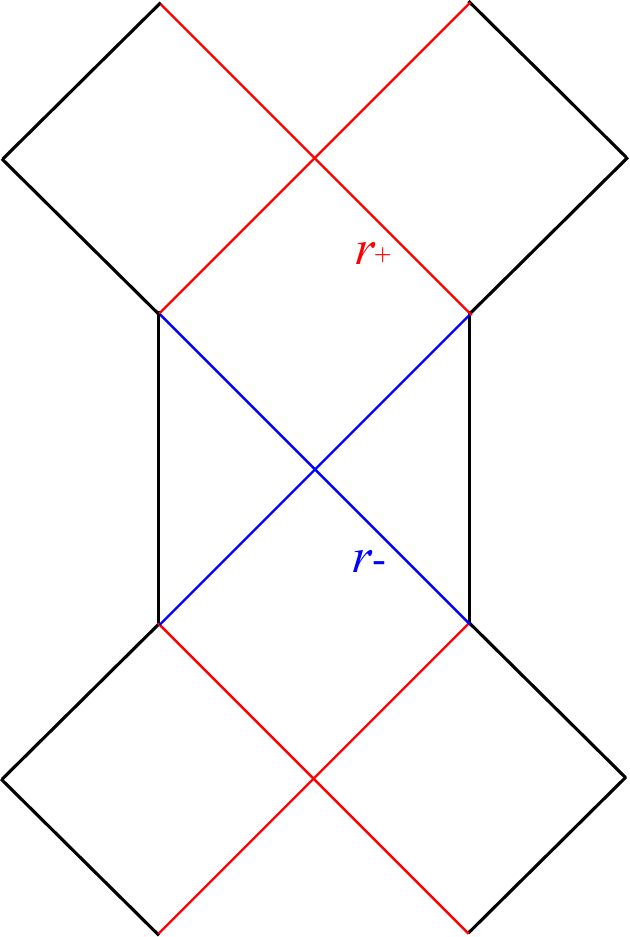}
\caption{\label{fig:pen0}The Penrose diagram of the model in Eq.~(\ref{eq:model}). The solution has two horizons, $r_{\pm}$, and the time-like center.}
\includegraphics[scale=0.6]{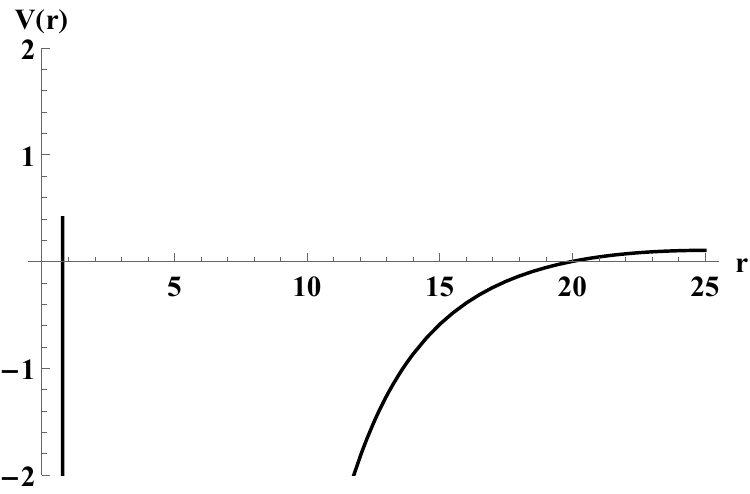}
\includegraphics[scale=0.6]{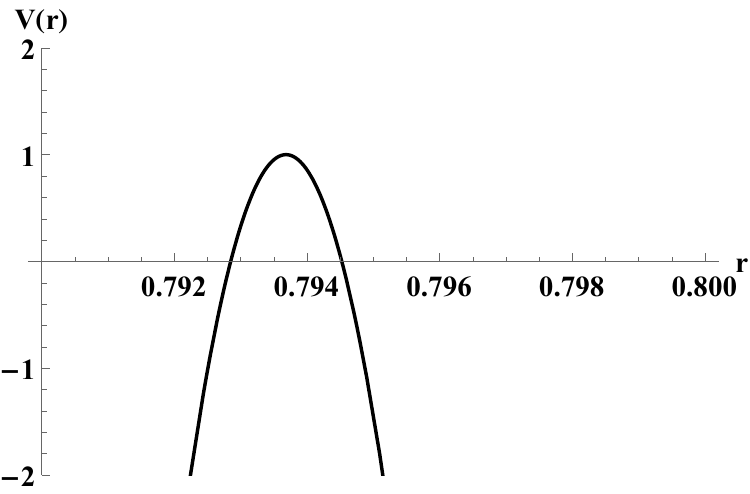}
\includegraphics[scale=0.6]{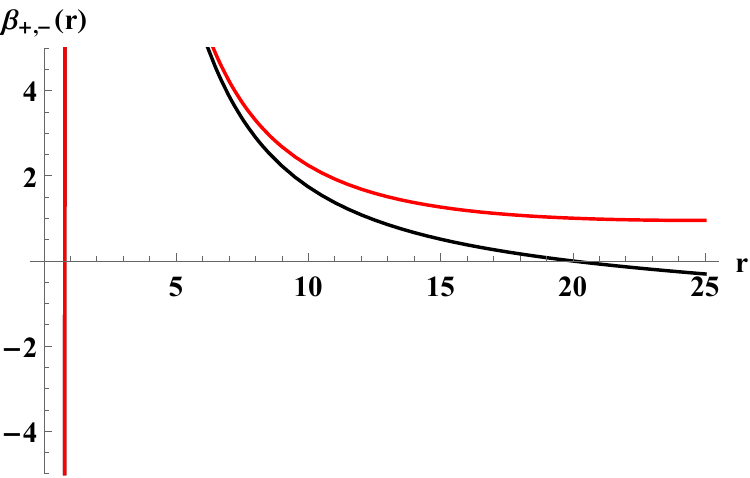}
\caption{\label{fig:tl}Dynamics of the time-like shell. Top left: $V_{\mathrm{eff}}$ with $M=10$, $A=0.1$, and $\sigma_{0} = 0.04$. There are two bouncing points located at $r_{\mathrm{max}} \simeq 19.9993$ and $r_{\mathrm{min}}\simeq 0.795$. Note that the outer horizon is $r_{+} = 19.9987$ and the inner horizon is $r_{-} = 0.8046$. Top right: $V_{\mathrm{eff}}$ around $r_{\mathrm{min}}\simeq 0.795$. Bottom: $\beta_{+}$ (black) and $\beta_{-}$ (red). This shows that for $r_{\mathrm{min}} \leq r \leq r_{\mathrm{max}}$, $\beta_{\pm} >0$ conditions are satisfied.}
\end{center}
\end{figure}

We consider the black hole model defined in\cite{Han:2023wxg}, which has a quantum-corrected center. The metric is
\begin{eqnarray}\label{eq:model}
ds^{2} = - f(r) dt^{2} + \frac{1}{f(r)} dr^{2} + r^{2} d\Omega^{2},
\end{eqnarray}
with
\begin{eqnarray} \label{fr}
f(r) = 1 - \frac{2M}{r} + \frac{A M^{2}}{r^{4}},
\end{eqnarray}
where $M$ is the black hole mass, while $A$ is a constant. Generically, this geometry has two horizons, labeled with $r_{\pm}$, with a time-like center (Fig.~\ref{fig:pen0}).
We note here that the bounce in this model is driven by the $AM^2/r^4$ term in Eq.~(\ref{fr}). This term 
dominates only at small values of $r$, and provides a repulsive gravity. Thus, directly from this form, we expect an oscillating behavior. At large values of $r$, the attractive term, $2M/r$, dominates and drives the collapse. At some minimal value of $r$, the repulsive term causes bounce and pushes the collapsing object out to larger values of $r$ where the attractive term again dominates and the cycle starts again.

\section{\label{sec:tim}Time-like thin-shells and gravitational collapses}

We first consider a thin time-like shell in order to explain the process of gravitational collapse in the framework of the model given in Eq.~(\ref{eq:model}).

\subsection{Junction equations}

The metric outside and inside the shell is
\begin{eqnarray}
ds_{\pm}^{2} = - f_{\pm}(r) dt^{2} + \frac{1}{f_{\pm}(r)} dr^{2} + r^{2} d\Omega^{2},
\end{eqnarray}
where $+$ and $-$ stand for outside and inside the shell, respectively. The metric of the time-like shell is
\begin{eqnarray}
ds_{\mathrm{shell}}^{2} = - d\tau^{2} + r^{2}(\tau) d\Omega^{2}.
\end{eqnarray}
Here, we assume $f_{+} = f(r)$ and $f_{-} = 1$.

After imposing the junction equation \cite{Israel:1966rt}, we obtain
\begin{eqnarray}
\epsilon_{-} \sqrt{\dot{r}^{2} + f_{-}} - \epsilon_{+} \sqrt{\dot{r}^{2} + f_{+}}  = 4\pi r \sigma(r),
\end{eqnarray}
where $\sigma(r)$ is the tension of the shell, and $\epsilon_{\pm} = \pm 1$ are the signs of the extrinsic curvatures. Here, extrinsic curvatures $\beta_{\pm}$ are
\begin{eqnarray}
\beta_{\pm} \equiv \frac{f_{-} - f_{+} \mp 16 \pi^{2} \sigma^{2} r^{2}}{8\pi \sigma r} = \epsilon_{\pm} \sqrt{\dot{r}^{2} + f_{\pm}}.
\end{eqnarray}
Note that if $\epsilon_{\pm} = +1$, $r$ increases along the outward normal direction, while if $\epsilon_{\pm} = -1$, $r$ decreases along the outward normal direction. Therefore, we have to assume $\epsilon_{\pm} = 1$.

After simple computations, we obtain the equation:
\begin{eqnarray}
\dot{r}^{2} + V_{\mathrm{eff}} (r) = 0,
\end{eqnarray}
where
\begin{eqnarray}
V_{\mathrm{eff}} (r) = f_{+} - \frac{\left(f_{-} - f_{+} - 16 \pi^{2} \sigma^{2} r^{2} \right)^2}{64 \pi^2 \sigma^2 r^2}.
\end{eqnarray}
Here, we interpret that $V_{\mathrm{eff}} < 0$ corresponds to the region where classical trajectories are allowed.

\subsection{Analysis of the solution}

In order to form a black hole, one can set $\sigma = \sigma_{0}$, and assume $\lambda/\sigma = -1$, where $\lambda$ is the pressure of the shell. For example, a scalar field can satisfy such a condition \cite{Blau:1986cw}.

Fig.~\ref{fig:tl} is an example that describes the gravitational collapse of a time-like shell and the formation of a black hole. Top left and right of Fig.~\ref{fig:tl} are $V_{\mathrm{eff}}$, where we choose $M=10$, $A=0.1$, and $\sigma_{0} = 0.04$. For these values of parameters, $r_{+} = 19.9987$ and $r_{-} = 0.8046$. By evaluating $V_{\mathrm{eff}}$, we find two bouncing points $r_{\mathrm{max}} \simeq 19.9993$ and $r_{\mathrm{min}}\simeq 0.795$. Therefore, $r_{\mathrm{min}} < r_{-}$ and $r_{\mathrm{max}} > r_{+}$, and hence, the shell propagates from the region outside of the outer horizon to the region inside of the inner horizon. In addition, bottom of Fig.~\ref{fig:tl} shows $\beta_{+}$ (black) and $\beta_{-}$ (red), which indicates that for a classically allowed region $r_{\mathrm{min}} \leq r \leq r_{\mathrm{max}}$, the extrinsic curvatures $\beta_{\pm}$ are always positive, as we expected.

\begin{figure}
\begin{center}
\includegraphics[scale=0.5]{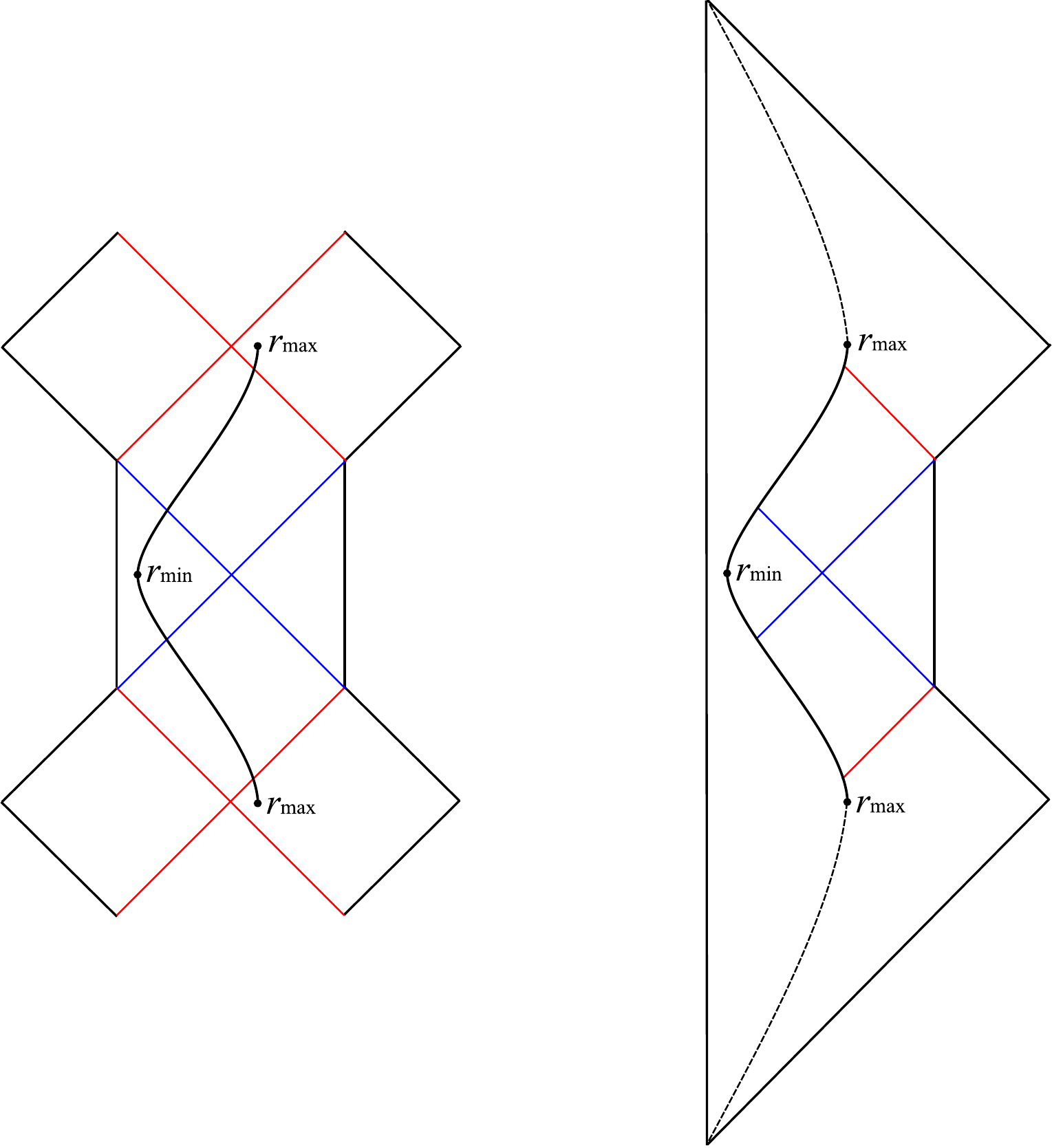}
\caption{\label{fig:pen1}Left: The Penrose diagram of the black hole solution, where the red lines are outer horizons and the blue lines are inner horizons. There exists a time-like shell solution that is oscillating between $r_{\mathrm{min}} \leq r \leq r_{\mathrm{max}}$, where $r_{\mathrm{max}}$ is outside the outer horizon and $r_{\mathrm{min}}$ is inside the inner horizon. Right: Inside the shell, the geometry is Minkowski. This diagram represents the resulting spacetime of the black hole formation.}
\end{center}
\end{figure}

If we summarize these numerical results, one can conceptually reconstruct Fig.~\ref{fig:pen1} as a Penrose diagram. The time-like shell is located between $r_{\mathrm{min}} \leq r \leq r_{\mathrm{max}}$, where $r_{\mathrm{max}}$ is outside the outer horizon and $r_{\mathrm{min}}$ is inside the inner horizon. Using the cut-and-paste technique, we paste a Minkowski space inside the shell. On the right side of the Fig.~\ref{fig:pen1}, there are dashed curves. These curves apparently do not follow the thin-shell trajectories. However, assuming some properties of a star interior, it is reasonable to assume that such a stationary shell is located outside the horizon \cite{Chen:2015lbp}.

\section{\label{sec:spa}Space-like thin-shells and black hole fireworks}

To consider the black hole firework scenario, we need to cut and paste on top of Fig.~\ref{fig:pen1}. We will introduce a space-like shell and use it to paste two space-like slices \cite{Brahma:2018cgr}.

\subsection{Junction equations}

The metric outside and inside the shell:
\begin{eqnarray}
ds_{\pm}^{2} = - \frac{1}{\tilde{f}_{\pm}(r)} dr^{2} + \tilde{f}_{\pm}(r) dt^{2} + r^{2} d\Omega^{2},
\end{eqnarray}
where $+$ and $-$ denote outside and inside the shell. The metric of the space-like shell is
\begin{eqnarray}
ds_{\mathrm{shell}}^{2} = ds^{2} + r^{2}(s) d\Omega^{2}.
\end{eqnarray}
Here, we impose that
\begin{eqnarray}
\tilde{f}_{\pm}(r) = - f(r) = - 1 + \frac{2M}{r} - \frac{A M^{2}}{r^{4}},
\end{eqnarray}
in other words, the regions outside and inside the shell correspond to the black hole solution in question.

\begin{figure}
\begin{center}
\includegraphics[scale=0.6]{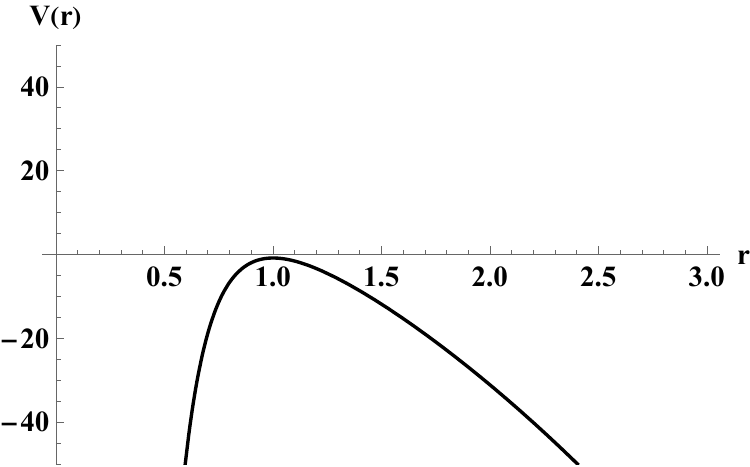}
\includegraphics[scale=0.6]{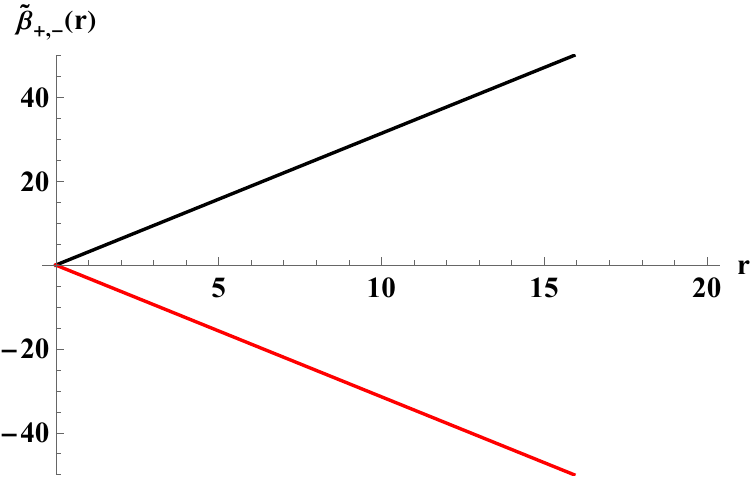}
\caption{\label{fig:sl}Dynamics of a space-like shell. Left: $V_{\mathrm{eff}}$ with $M=10$, $A=0.1$, $\sigma_{0} = -0.5$, and $w = -1$. This shows that the space-like shell covers the space from infinity to the center. Right: $\tilde{\beta}_{+}$ (black) and $\tilde{\beta}_{-}$ (red). This shows that  $\tilde{\beta}_{+} >0$ and $\tilde{\beta}_{-} < 0$ as expected.}
\includegraphics[scale=0.6]{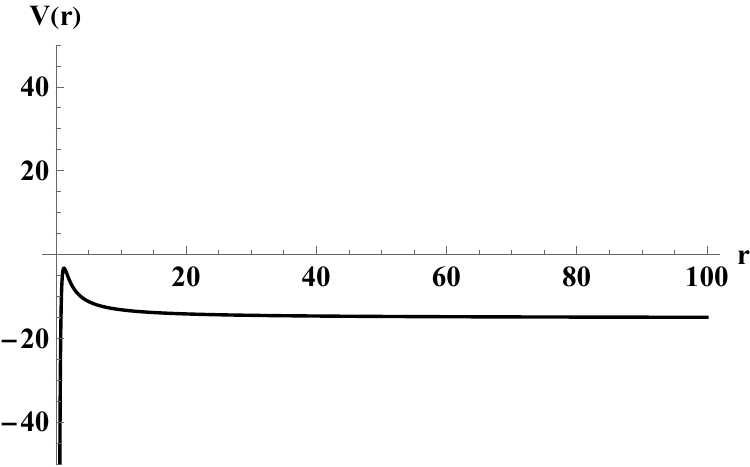}
\includegraphics[scale=0.6]{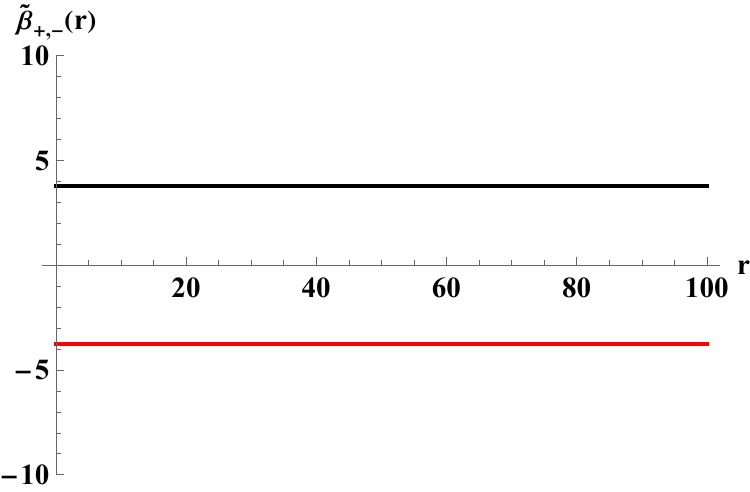}
\caption{\label{fig:sl34}Another example of a space-like shell. Left: $V_{\mathrm{eff}}$ with $M=10$, $A=0.1$, $\sigma_{0} = -0.6$, and $w = -0.5$. Again, the space-like shell covers the region from infinity to the center. Right: $\tilde{\beta}_{+}$ (black) and $\tilde{\beta}_{-}$ (red). This shows that $\tilde{\beta}_{+} >0$ and $\tilde{\beta}_{-} < 0$ as expected.}
\end{center}
\end{figure}

After imposing the junction equation \cite{Balbinot:1990zz}, the result is
\begin{eqnarray}
\epsilon_{-} \sqrt{\dot{r}^{2} + \tilde{f}_{-}} - \epsilon_{+} \sqrt{\dot{r}^{2} + \tilde{f}_{+}}  = 4\pi r \sigma(r),
\end{eqnarray}
where $\sigma(r)$ is the tension of the shell, and $\epsilon_{\pm} = \pm 1$ are the signs of the extrinsic curvatures. Here, the extrinsic curvatures $\tilde{\beta}_{\pm}$ are
\begin{eqnarray}
\tilde{\beta}_{\pm} \equiv \frac{\tilde{f}_{-} - \tilde{f}_{+} \mp 16 \pi^{2} \sigma^{2} r^{2}}{8\pi \sigma r} = \epsilon_{\pm} \sqrt{\dot{r}^{2} + \tilde{f}_{\pm}}.
\end{eqnarray}
Note that if $\epsilon_{\pm} = +1$, $r$ increases along the outward normal direction (direction from future to the past), while if $\epsilon_{\pm} = -1$, $r$ decreases along the outward normal direction. Therefore, in our case, we assume that $\epsilon_{+} = +1$ and $\epsilon_{-} = -1$. Hence, $\sigma < 0$ is required, and the null energy condition must be violated. This is expected because of the repulsive term in Eq.~(\ref{fr}).

After simple computations, we obtain the equation
\begin{eqnarray}
\dot{r}^{2} + \tilde{V}_{\mathrm{eff}} (r) = 0,
\end{eqnarray}
where
\begin{eqnarray}
\tilde{V}_{\mathrm{eff}} (r) = \tilde{f}_{+} - \frac{\left(\tilde{f}_{-} - \tilde{f}_{+} - 16 \pi^{2} \sigma^{2} r^{2} \right)^2}{64 \pi^2 \sigma^2 r^2}.
\end{eqnarray}

\begin{figure}
\begin{center}
\includegraphics[scale=0.5]{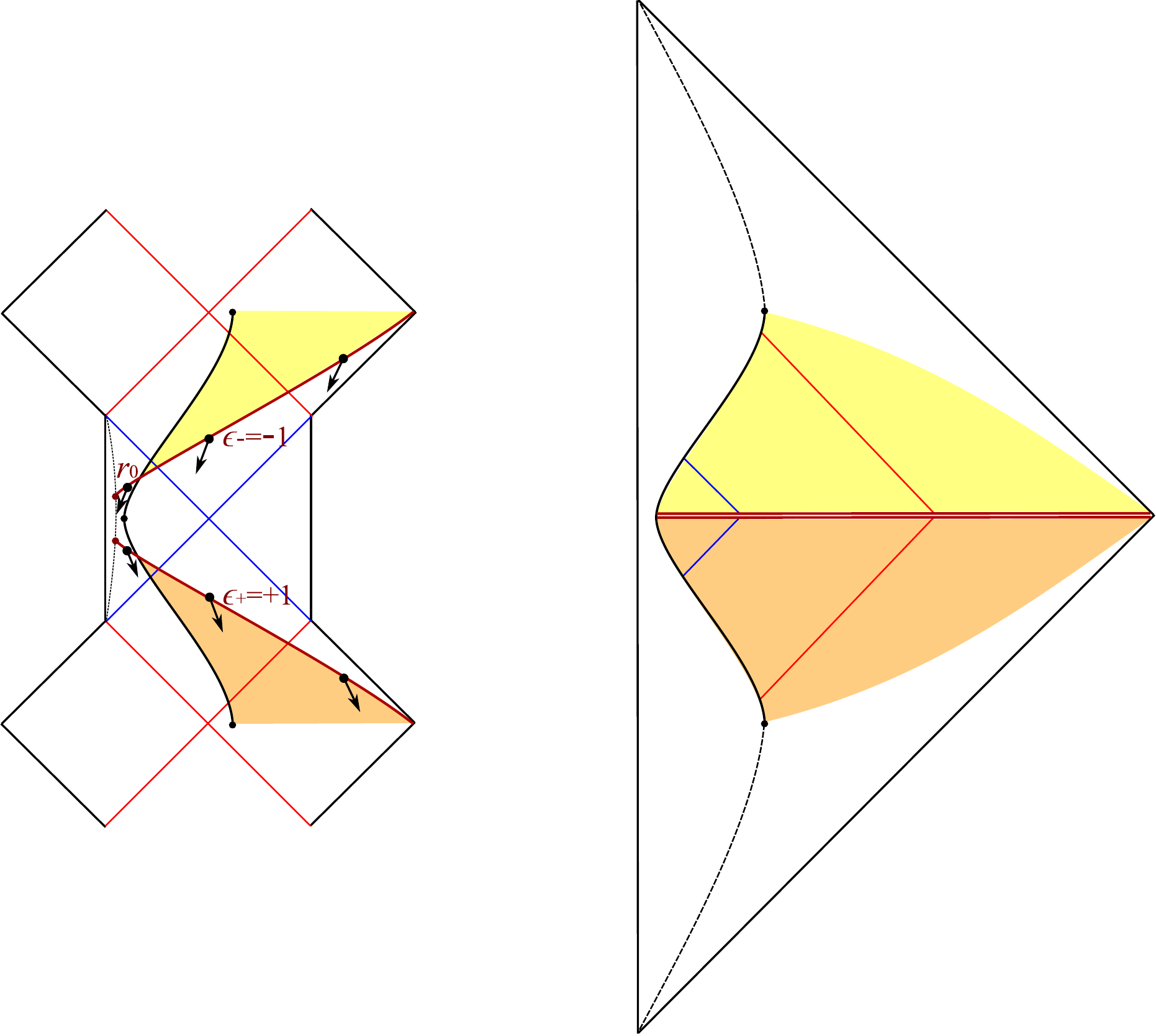}
\caption{\label{fig:pen2}Left: The space-likes shells with $\epsilon_{-} = -1$ (upper) and $\epsilon_{+} = +1$ (lower), where small black arrows denote the outward normal direction. We paste the future of the upper shell ($\tilde{f}_{-}$, yellow-colored region) and the past of the lower shell ($\tilde{f}_{+}$, orange-colored region). Right: After we paste two regions, we obtain the final causal structure of the black hole fireworks.}
\end{center}
\end{figure}

\subsection{Conditions for thin-shells}

We now need to assume the condition for the thin-shell. The energy conservation equation is
\begin{eqnarray}
\dot{\sigma} = -2 \frac{\dot{r}}{r} \left( \sigma - \lambda \right),
\end{eqnarray}
where $\lambda$ is the pressure of the shell. If we assume the equation of state of the space-like shell $w_{i} = - \lambda_{i}/\sigma_{i}$ to be a constant, the generic solution of this equation is
\begin{eqnarray}
\sigma(r) = \sum_{i} \frac{\sigma_{0i}}{r^{2(1+w_{i})}},
\end{eqnarray}
where $\sigma_{0i}$ are constants.

By assuming a specific function of the tension, we want to impose the following conditions:
\begin{itemize}
\item[--] 1. The shell covers the region from $r_{0} < r_{\mathrm{min}}$ to infinity, i.e., $\tilde{V}(r) < 0$ for $r_{0} \leq r \leq \infty$.
\item[--] 2. Extrinsic curvatures satisfy $\tilde{\beta}_{+} > 0$ and $\tilde{\beta}_{-} < 0$ for $r_{\mathrm{min}} \leq r \leq \infty$.
\end{itemize}

\subsection{Construction of the black hole firework geometry}

In order to satisfy the extrinsic curvature conditions, the null energy condition of the shell must be violated. For example, Fig.~\ref{fig:sl} shows the case when the shell has a constant negative tension (a domain wall case). Fig.~\ref{fig:sl34} shows the case where the tension asymptotically approaches zero at infinity ($w = -0.5$ and $\sigma \sim 1/r$), and thus the negative tension effects disappear at infinity.

After we cut and paste the spacetimes outside and inside the shell, we obtain the causal structure in Fig.~\ref{fig:pen2}. Outside the shell satisfies $\epsilon_{-} = -1$, while inside the shell satisfies $\epsilon_{+} = +1$. We paste the future of the outer shell (yellow-colored region) and the past of the inner shell (orange-colored region). As a result, we obtain the final causal structure of the black hole fireworks (right of Fig.~\ref{fig:pen2}).

Using the thin-shell approximation, it is possible to justify such a causal structure. If $r_{0} \neq r_{\mathrm{min}}$, the time-like and space-like shell can intersect. A complete description of this intersection perhaps belongs to the regime of quantum gravity, however it is still interesting to ask what happens there in the framework of general relativity.

The only price that we need to pay for this construction is a violation of the null energy condition, even outside of the horizon \cite{Brahma:2018cgr}. In addition, we need to ask whether the causal structure of the analytic solution will still be valid in dynamical situations. The inner horizon might be unstable due to the mass inflation \cite{Poisson:1990eh}. If we take this into account, we may not be able to trust the causal structure inside the event horizon.

\section{\label{sec:bou}Bouncing time-scale for black hole firework scenarios}

In this section, we discuss the bouncing time observed by different observers. We use the same assumption that the quantum gravity corrections should be small, \textit{i.e.} $A \sim m^2_{Pl} \ll M^2$. In this limit, the parameter $A$ plays no role in the leading order estimate, and therefore, the mundane Schwarzschild solution is sufficient for this particular discussion.\footnote{Though, we comment on the possible relation between $\delta$ and $A$ later in Sec.~\ref{SubSec:BounceT_discussion}.} We first re-derive the bouncing time for a distant observer at a fixed $R$ as presented in Ref.~\cite{Han:2023wxg}. After that, we calculate the bouncing time measured by the observer who is comoving with the shell, which we believe is more relevant for the fireworks scenario.  We discuss the slicing dependence of the two different bouncing times in Sec.~\ref{SubSec:BounceT_discussion}.

\subsection{Bouncing time scale with the $\delta$ parameter}\label{Subsec.Bouncing time delta}

\begin{figure}
\begin{center}
\includegraphics[scale=0.35]{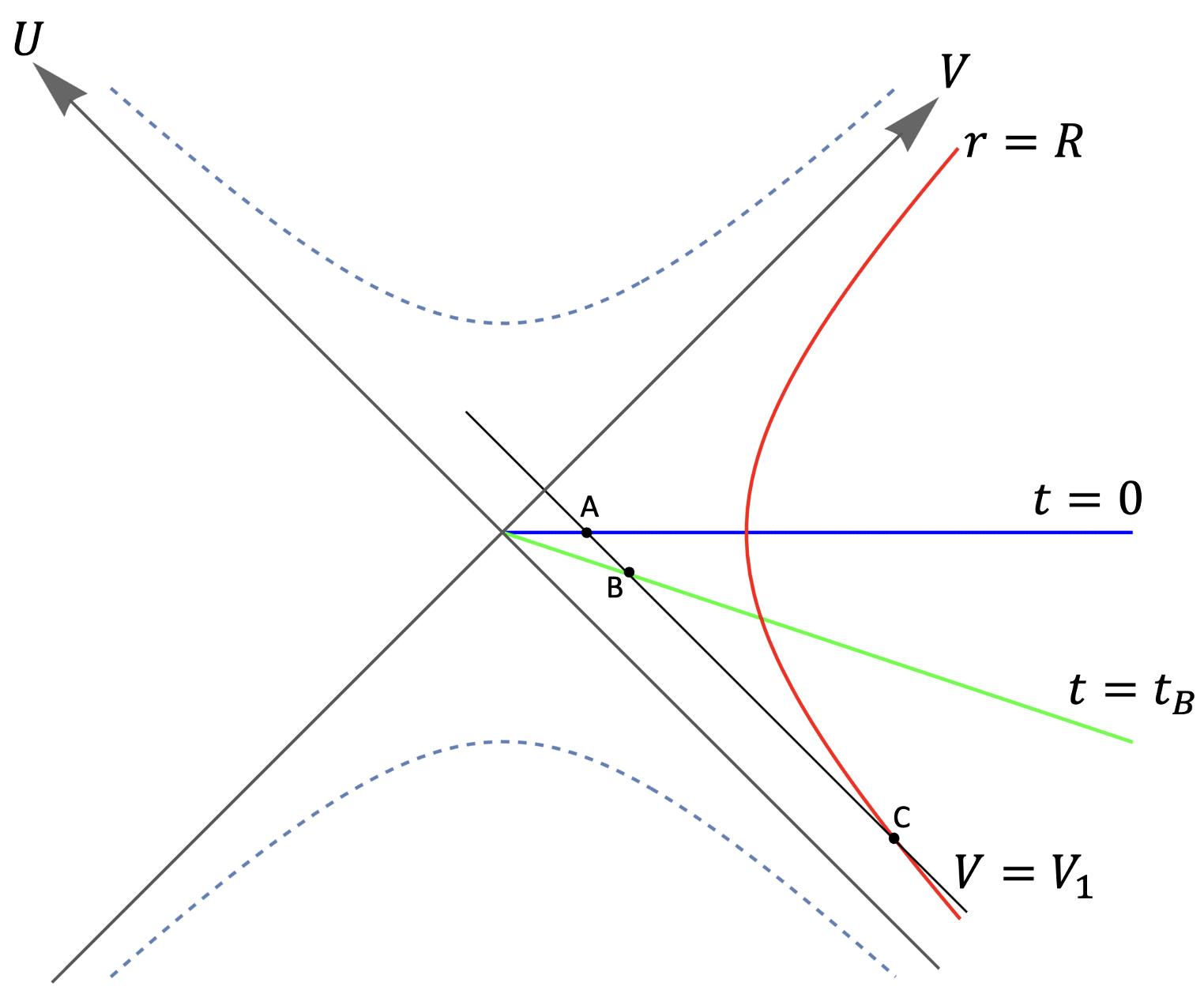}
\caption{\label{fig:bouncing_time}Bouncing time in the Schwarzschild metric. In Ref.~\cite{Han:2023wxg}, the trajectory of $V=V_{1}$ is uniquely determined by the corresponding spacetime diagram therein. However, the bouncing time, defined by $\mathcal{T} \equiv -4M \ln \delta $, can be shown to be from a geometric relation in the Schwarzschild spacetime, which is related to the arbitrary cutting of spacetime. }
\end{center}
\end{figure}

Outside the horizon, the black hole is well approximated by the Schwarzschild solution. Assuming the Schwarzschild solution, the double-null coordinates $U$ and $V$ satisfy
\begin{eqnarray}\label{KS_UV}
    UV=\left(1-\frac{r}{2M}\right)e^{r/2M}
\end{eqnarray}
and
\begin{eqnarray}\label{KS_U/V}
  \frac{U}{V}=-e^{-t/2M},   
\end{eqnarray}
in the region of interest (see Fig.~\ref{fig:bouncing_time}).
In Fig.~\ref{fig:bouncing_time}, events A, B, and C are given by the intersections of a specific ingoing light ray $V=V_{1}$ with the constant-$t$ hypersurfaces, $t=0$ and $t=t_{B}$, and a constant-$r$ trajectory $r=R$, respectively.
Since $t_{B}$ is an arbitrary constant, we can also consider an event A to be a special case where we consider $t_{B}=0$.  

In terms of the Schwarzschild coordinate $(t, r)$, the locations of B and C are given by $(t_B, 2M+\Delta)$ and $(t_C, R)$, respectively. 
Using (\ref{KS_UV}) and (\ref{KS_U/V}), one can show that the quantities $\Delta$, $t_B$, and $t_C$ satisfy the following relation
\begin{equation}
 \frac{\left(\frac{R}{2M}-1\right)e^{R/2M}}{\frac{\Delta}{2M}e^{(1+\Delta/2M)}}=e^{\tilde{T}/4M}, 
\end{equation}
where $\tilde{T} \equiv 2(t_B-t_C)$. Assuming $\Delta \ll 2M \ll R$, the above relation reduces to 
\begin{equation}\label{general_R_Btime}
\tilde{T} \approx 2R +4M \ln R -4M \ln \Delta.
\end{equation}
By choosing $t_B=0$, \textit{i.e.} considering event A with location $(t, r)=(0, 2M+\delta)$, we obtain the relation given in Ref.~\cite{Han:2023wxg} 
\begin{equation}
T \approx 2R +4M \ln R -4M \ln \delta,
\end{equation}
where $T/2 \approx -t_C$. Since we used the relation $2M \gg R$, this interval of the Schwarzschild time $T$ is approximately the bouncing time measured by a distant observer fixed at $r=R$. Eq.~(\ref{general_R_Btime}) has a simple physical interpretation.  A distant observer shoots a ray of light radially into the black hole (event C). After $\tilde{T}/2$ of this observer's proper time elapsed, he/she would think the light ray is $\Delta$ away from the event horizon. 

\subsection{Bouncing time for the comoving observer}\label{SubSec:BounceT_comoving}



Apart from the asymptotic observer whose coordinate system is incomplete, there is another observer who is perhaps more relevant to the bouncing process. This is an observer comoving with the collapsing shell. Thus, a more appropriate physical time scale can be calculated using the proper time of the observer that crosses the event horizon. One can easily evaluate the proper time of the time-like shell that transitions from the black hole to the white hole phase as
\begin{equation}
\tau = 2 \left| \int_{R}^{r_{\mathrm{min}}} \frac{dr}{\sqrt{-V(r)}} \right|.
\end{equation}
Here we use a collapsing shell of (pressureless) dust as a demonstration. In this case, the rest mass of the dust $\alpha$ is assumed to be conserved and is given by $\alpha=4\pi r^2 \sigma =const.$. From the Israel junction conditions, we obtain
\begin{equation}
M=\alpha \sqrt{1+\dot{r}^2}-\frac{\alpha^2}{2r},
\end{equation}
where the overdot is the derivative with respect to the proper time along the timelike trajectory of the infalling shell. From this, one can compute the proper time elapsed along the shell trajectory for one complete cycle as follows
\begin{equation}
 \tau = 2 \left| \int_{R_{\mathrm{max}}}^{r_{\mathrm{min}}} \frac{dr}{\sqrt{\left(\frac{M}{\alpha}+\frac{\alpha}{2r}\right)^2-1}} \right|. 
\end{equation}
To have $R_{\mathrm{max}}$ finite, \textit{i.e.} the shell is bounded, we must have $\alpha > M$, for which $R_{\mathrm{max}}= \frac{\alpha^2}{2(\alpha-M)}$. In this case, the above integral is given by
\begin{equation}\label{tau_exact}
 \tau=\left. 2\sqrt{\frac{1}{1-\frac{m^2}{\alpha ^2}}} \left(\frac{1}{2} C
   \tan ^{-1}\left(\frac{2 r-C}{2 \sqrt{C
   r+B-r^2}}\right)-\sqrt{C r+B-r^2}\right) \right\vert_{r_{\mathrm{min}}}^{R_{\mathrm{max}}},   
\end{equation}
where $C=\frac{m}{1-\frac{m^2}{\alpha ^2}}$ and $B=\frac{\alpha ^2}{4 \left(1-\frac{m^2}{\alpha ^2}\right)}$.
If we further consider large $R_{\mathrm{max}} \gg M$, based on the relation for $R_{\mathrm{max}}$, we also have $\alpha \sim 2R_{\mathrm{max}}$. Thus, the above integration is approximately given by 
\begin{equation}\label{tau_approx}
 \tau \sim 2 \left| \int_{R_{\mathrm{max}}}^{r_{\mathrm{min}}} \frac{dr}{\sqrt{\left(\frac{\alpha}{2r}\right)^2-1}} \right|\sim 2 \left| \int_{R_{\mathrm{max}}}^{r_{\mathrm{min}}} \frac{dr}{\sqrt{\left(\frac{R_{\mathrm{max}}}{r}\right)^2-1}} \right|= 2\sqrt{R^2_{\mathrm{max}}-r^2_{\mathrm{min}}}.  
\end{equation}
In this limit, the bouncing time is mostly determined by $R_{\mathrm{max}}$, while the exact value of $r_{\mathrm{min}}$ is not that important. This is physically reasonable since in this limit, the shell's velocity relative to the center of the black hole is high when $r$ is small.  
If we include the quantum gravity modification, \textit{i.e.} using the metric in Eq.~(\ref{fr}) instead of the Schwarzschild solution, the shell will be repelled at some minimal radius due to the extra repulsive term $AM^2/r^4$. Thus, we have to set the $r_{\mathrm{min}}$ to be the bouncing point inside the horizon, which is determined by the three parameters $\{\alpha, M, A\}$. However, due to the smallness of $AM^2$, the bouncing point must be deep inside the event horizon and the modifications to Eqs.~(\ref{tau_exact}) and (\ref{tau_approx}) are small.

\subsection{Interpretation: coordinate time and proper time}\label{SubSec:BounceT_discussion}

\begin{figure}
\begin{center}
\includegraphics[scale=0.5]{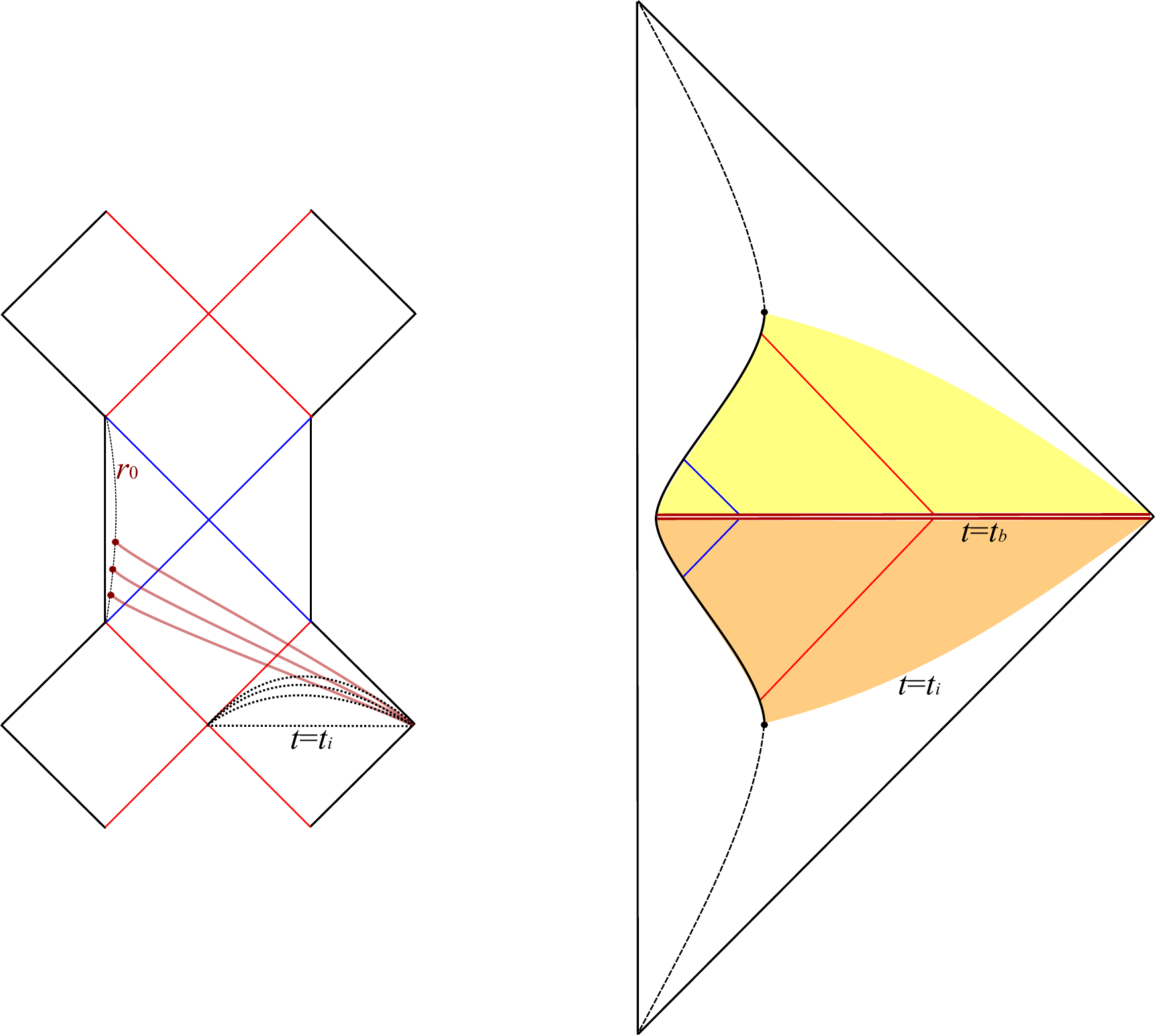}
\caption{\label{fig:pen3}Left: Due to the time-translation symmetry, there can be several equivalent spacelike slices (red curves) that have a different coordinate time at infinity. Black dotted curves correspond to constant $t$ hypersurfaces. Right: The bouncing time is the difference between $t = t_{i}$ and $t = t_{b}$, where $t_{b}$ is arbitrary.}
\end{center}
\end{figure}

We now analyze the coordinate time difference between two slices (Fig.~\ref{fig:pen3}). Due to the \textit{time-translation symmetry}, one can choose an arbitrary coordinate time (at infinity) for the space-like hypersurface (left of Fig.~\ref{fig:pen3}). This means that the time difference between the $t=t_{i}$ (that can be chosen in the sufficient past) and $t=t_{b}$ (the bouncing time inside the horizon) is arbitrary in this setup (right of Fig.~\ref{fig:pen3}). 
According to the discussion in Sec.~\ref{Subsec.Bouncing time delta}, one may find a corresponding $\delta$ parameter to denote the $t=t_{b}$ hypersurface. As already mentioned in Ref.~\cite{Han:2023wxg}, the bouncing time for the distant observer is determined by how one cuts and pastes the spacetime outside the event horizon. The same argument is valid for the spacelike slicing considered in our case  (see Fig.~\ref{fig:pen3}). Therefore, this parameter $\delta$ is not really appropriate to parameterize the physical bouncing time. On the other hand, the bouncing time measured by the comoving observer discussed in Sec.~\ref{SubSec:BounceT_comoving} is very different in this aspect. Based on the previous discussion, the contribution to the bouncing time around $r_{\mathrm{min}}$ is small, so even if we cut out a certain portion of the spacetime as in Fig.~\ref{fig:pen2}, the corresponding proper time is only mildly affected by the cut-and-paste procedure. Interestingly, this bouncing time can be unambiguously determined in the model considered in Ref.~\cite{Han:2023wxg} since the trajectory of the shell (or surface of the collapsing star) is intact by the designed cut (see Fig.~4 therein). One can easily construct the scenario in which two observers (the comoving and the fixed-$r$ one) begin their journeys at the same spacetime event when the shell is at some $R_{\mathrm{max}}$. After the whole period of the bounce, in the absence of any dissipation (as in Refs.~\cite{Haggard:2014rza} and ~\cite{Han:2023wxg}), the two observers meet each other again at the next $R_{\mathrm{max}}$ defined for the comoving observer. It is possible that some relation exists between the parameter $A$, which modifies the proper time of the comoving observer due to the quantum gravity correction to the Schwarzschild solution, and the parameter $\delta$, which controls the proper time of the observer fixed at $R_{\mathrm{max}}$. 


Finally, in addition to them, indeed, we need to include Hawking radiation and its back-reaction. To describe smoothly the evolution of the collapsing object from infinity to black hole horizons, we need to study a dynamical causal structure of the spacetime from formation to evaporation including the back-reaction of the background geometry. When a black hole model has two horizons that disappear within a finite proper time, i.e. an apparent horizon has a circular shape in the Penrose diagram, then there should be a smooth way to describe the entire spacetime without any divergences; for example, see \cite{Hwang:2012nn,Brahma:2019oal}. Applications to the present model are left for a future research project.

\section{\label{sec:dis}Discussion} 

In this paper, we revisited some aspects of the black hole fireworks (i.e. a black hole to white hole transition) scenario proposed in \cite{Haggard:2014rza,Han:2023wxg}. We constructed an explicit model for the black hole fireworks using the cut-and-paste technique. First, we used evolution of a time-like shell in the background of the loop quantum gravity inspired metric to model the process of gravitational collapse. Then using the space-like shell analysis, we constructed the firework geometry. We used the well defined thin-shell techniques where all the relevant quantities are clearly defined. Thus, our analysis removes some subtle issues that were present in the previous literature. 

We showed that the firework scenario requires specific conditions outside the event horizon, in principle the violation of the energy conditions. This can be expressed in terms of the tension of the space-like junction where the two metrics meet.  In particular, we used a rather simple and well-studied space-like junction technique to create the black to white hole bounce with a single asymptotic region. For comparison, in Ref.~\cite{Han:2023wxg}, a more complicated cut-and-paste procedure is utilized to achieve the same goal, without violating the null energy condition away from the horizon.  However, such a cut corresponds to a hypersurface which changes its characteristic from spacelike to null. The tension conditions for such a scenario are highly non-trivial and might not be physically justifiable. We leave this issue for future work.

We also calculated the proper and coordinate time scales required for the black hole to white hole transition. The proper time scale is classical and hence it must be sufficiently shorter than the evaporation time scale. However, we point out that the coordinate time scale (related to the $\delta$-parameter in the black hole firework scenario in  \cite{Han:2023wxg}) can be chosen arbitrarily. 
The bouncing time for the distant observer is determined by how one cuts and pastes the spacetimes outside the event horizon, and thus does not have any obvious connection to quantum gravity effects.


\newpage

\section*{Acknowledgment}

DY and WL was supported by the National Research Foundation of Korea (Grant No. : 2021R1C1C1008622, 2021R1A4A5031460). DS is partially supported by the US National Science Foundation, under Grants No. PHY-2014021 and PHY-2310363.

\end{document}